\documentclass[10pt,letterpaper]{article}

\usepackage{cogsci}
\usepackage{graphicx}
\newcommand{\citet}[1]{ \shortcite{#1}}
\newcommand{\citep}[1]{\citeyear{#1}}
\newcommand{\citealp}[1]{\citeauthor*{#1} \citeyear{#1}}

\cogscifinalcopy 

\usepackage{pslatex}
\usepackage{apacite}
\usepackage[symbol]{footmisc}

\begin{document}

\author{{\large \bf Lance Ying (lancelcy@umich.edu)} \\
  Department of EECS, University of Michigan
  \AND {\large \bf Audrey Michal (almichal@umich.edu)} \\
  Department of Psychology, University of Michigan
  \AND {\large \bf Jun Zhang (junz@umich.edu)\thanks{Corresponding author}} \\
  Department of Psychology and Department of Statistics, University of Michigan}
\title{A Bayesian Drift-Diffusion Model of Schachter-Singer's Two Factor Theory of Emotion}

\maketitle

\begin{abstract}
Bayesian inference has been used in the past to model visual perception \cite{kersten2004object}, accounting for the Helmholtz principle of perception as “unconscious inference” that is constrained by bottom-up sensory evidence (likelihood) while subject to top-down expectation, priming, or other contextual influences (prior bias); here "unconsciousness" merely relates to the "directness" of perception in the sense of Gibson. Here, we adopt the same Bayesian framework to model emotion process in accordance with Schachter-Singer’s Two-Factor theory, which argues that emotion is the outcome of cognitive labeling or attribution of a diffuse pattern of autonomic arousal (Schachter \& Singer, 1962). In analogous to visual perception, we conceptualize the emotion process, in which emotional labels are constructed, as an instance of Bayesian inference, either consciously or unconsciously combining the contextual information with a person’s physiological arousal patterns. Drift-diffusion models were constructed to simulate emotional processes, where the decision boundaries correspond to the emotional state experienced by the participants, and boundary-crossing constitutes “labeling” in Schachter-Singer’s sense. Our model is tested against experimental data from the Schachter \& Singer's study (1962) on context-modulated emotional state labeling and the Ross et al. study (1969) on fear reduction through mis-attribution. Two model scenarios are investigated, in which arousal pattern as one factor is pitted against contextual interaction with an confederate (in Schachter-Singer case) or explicitly instructed mis-attribution (in Ross et al. case) as another factor, mapping onto the Bayesian prior (initial position of the drift) and the  likelihood function (evidence accumulation or drift rate). 
We find that the first scenario (arousal as the prior and context as the likelihood) has a better fit with Schachter \& Singer (1962) whereas the second scenario (context as the prior and arousal as the likelihood) has a better fit with Ross et al. (1969).


\end{abstract}

\section{Introduction}
In his "The Principles of Psychology", William James (1922) famously asked "what is an emotion", and focused on physiological arousal as a basis of emotional experience. This arousal theory of emotion subsequently inspired a huge volume of work on cognitive involvement, such as misattribution, appraisal, etc. \cite{russell2003core}. On the computation modeling side, the Bayesian approach to cognition has gained significant attention in recent years due to its capacity to solve induction and causal inference problems within a probabilistic framework \cite{griffiths2007two,l2008bayesian}, extending previous applications of Bayesianism to object perception \cite{kersten2004object} and language acquisition \cite{chater2006probabilistic}. A growing interest to apply Bayesian modeling to human emotions emerges, especially within the context of Theory of Mind \citet{saxe2017formalizing,ong2019computational}, though most existing works focus on third-person appraisals based on contextual cues instead of on identifying one's own emotion process.

In this paper, after a review of existing emotion theories, we propose a Bayesian formulation that computationally implements the Schachter \& Singer's Two-Factor theory of emotion. We focus on the Two-Factor theory as a conceptually viable framework for emotion computation for two reasons. First, the Two-Factor theory bridges the physiological and cognitive aspects of the emotion process \cite{reisenzein1983schachter}, with a variety of modern appraisal theories focusing on cognitive aspects. Second, we can demonstrate a clear mapping from the two factors in the Two-Factor theory, namely physiological arousal and cognitive label, to the two key components of the Bayesian inference scheme, the prior and the likelihood. This allows us to model first-person emotion recognition as a Bayesian inference process.

We then develop a drift-diffusion model to implement the dynamical Bayesian inference and account for experimental findings in Schachter \& Singer’s study. There, participants who were physiologically aroused (via drug injection but were not informed of arousal) later reported different emotions (i.e., labeled their arousal pattern differently) based on the nature of their interaction with an experimental confederate they encountered post-injection. In our drift-diffusion modeling, the decision boundaries correspond to the euphoric and anger state experienced by the participants in the experiment, and boundary-crossing constitutes “labeling” in Schachter-Singer’s sense. Response time (RT) in the drift-diffusion model is used as a surrogate measure of the self-rated intensity of the emotional state, where high intensity corresponds to a shorter response time. We propose two model scenarios (versions). In the first scenario, arousal pattern is used as the prior and the likelihood function for evidence accumulation models the interaction with the confederate (context). We adopt an unbiased prior, while allowing the drift-rate (and its sign) to capture the nature of interaction with the confederate. In the second scenario, we use the context as the prior and physiological arousal patterns as the likelihood function. The drift-diffusion model is then applied to account for the data of Ross et al. (1969), in which the time-course of the mis-attribution effect was empirically measured. Because the Ross et al. paradigm is one of decision-making under time pressure, we compare it to computation models of decision making with collapsing boundary. Simulation results for both the Schachter \& Singer's study (1962) on context-modulated emotional state labeling and the Ross et al. study (1969) on fear reduction through mis-attribution are presented.

\subsection{The Bayesian Approach to Perception}

\subsubsection{Bayes Formula}

Bayesian models have been increasingly used in cognition for decision making and prediction tasks. Bayesian inferences are based on the simple Bayes rule
 \begin{equation}
     P(h|e)=\frac{P(e|h)P(h)}{P(e)} = \frac{P(e|h)P(h)}{\sum_h P(e|h)P(h)}
 \end{equation}
 where $e$ represents an event and $h$ a hypothesis. Here $P(h)$ is called the prior, $P(e|h)$ is the likelihood function, and the resulting $P(h|e)$ is the posterior. In the Bayesian framework, uncertainty in reasoning about variables is reflected by a probability distribution, which is updated upon receiving evidence using probability theory \cite{lee2011cognitive}. The prior (and posterior) refers to the degree of belief in a hypothesis before (and after) observation, while the likelihood function represents the evidence accumulation process. The Bayesian approach has been extensively used in cognitive modeling as it allows agents to update knowledge about the world in a rational way (i.e., with self-consistency) after new data becomes available.

\subsubsection{Bayesian Inference and Perception}

It was Helmholtz who famously drew a distinction between sensation and perception, arguing that perception is an "unconscious inference" process based on sensory stimuli -- when sensory stimuli are registered, the mind draws inference on the reality where experience, context, and expectation play an important role \cite{westheimer2008helmholtz}.

Although Helmholtz's arguments are mostly made in a philosophical context, Bayesian inference was advanced as a computational framework in the study of how the brain can extract 3-D geometric information (e.g., shape) of objects, in a probabilistic way yet with an extraordinary level of accuracy, from 2-D visual inputs on the retina \cite{kersten2004object}. Based on these early successes in Bayesian modeling, emotion theorists have also attempted to apply similar frameworks on emotion perception (e.g. \cite{ong2019computational}).

\subsection{Theories and Paradigms of Emotion}

\subsubsection{Earlier theories}

Since the dawn of psychological science, scholars have proposed various models of emotional processes. For example, the James-Lange theory views emotion as the result of physiological arousal \cite{james1922emotions}, or as James puts it, "the feeling of bodily changes as they occur is the emotion". Therefore, emotion is felt when changes in bodily state are perceived. However, people who have lost sensations can still feel emotion and people who feel increased heart rate through exercise may not feel any particular emotion. Thus, physical sensation and emotion are clearly two different processes \cite{shouse2005feeling}. This essentially is the Cannon–Bard theory, arguing that physiological reactions and emotional experiences occur simultaneously, but are two independent processes \cite{cannon1927james}. Although the Cannon-Bard theory addressed some of the shortcomings of the James-Lange theory, it is challenged by studies showing that physical reactions can influence emotions \cite{coles2019meta}. 

\subsubsection{Two-Factor theory and its critique} 
An influential synthesis of these earlier conceptualizations is Schachter \& Singer's Two-Factor theory of emotion, which posits that emotion is based on two factors: physiological arousal and cognitive label \cite{schachter1962cognitive}. When a subject experiences arousal, the subject appraises the context of the arousal patterns, which leads to an experienced emotion state. In comparison to the James-Lange theory, the Two-Factor theory incorporates a cognitive component in the emotional process. In comparison to the Cannon-Bard theory, the intermediary role of cognition is established, i.e., mediating between physiological reaction and emotional experience and, according to Schachter \& Singer (1962), the presence of both autonomic arousal and a cognitive label are necessary for a person to perceive an emotion. The Two-Factor theory is supported by some studies on misattributed arousal and the theory is able to explain the emotion process in many situations. The theory also inspired many more modern variants of the cognition-arousal theory of emotion, which share the central postulate as the Two-Factor theory that emotion is a function of cognition and arousal, although they largely disagree on the ways in which cognition and arousal interact to generate emotion \cite{reisenzein2017varieties, lindquist2008constructing}.

However, the necessity of arousal for emotion is still a matter of intense debate \cite{reisenzein1983schachter}. A variant of the Two-Factor theory is the Cognitive Appraisal theory \cite{arnold1960emotion,lazarus1991emotion}, which also posits that emotions are felt due to the appraisals of the situations. Contrary to the Two-Factor account, Lazarus (1966) argues that cognitive appraisal {\it precedes} emotion and physiological arousal. Two types of appraisal were differentiated: in the primary appraisal, the subject appraises the relevance of the situation while in secondary appraisal, the subject evaluates the relevant resources for coping. Although supported by contemporary emotion research \cite{ellsworth2013appraisal,smith2014differentiation}, Cognitive Appraisal theory has been criticised for overemphasising the conscious or voluntary processes, as cognitive appraisal might only be one of several ways to produce emotion \cite{zajonc1980feeling,izard1993four}.

While the Two-Factor theory has sparked enormous research interest, including the Cognitive Appraisal theory, the experimental paradigm itself in the Schachter \& Singer famous study \cite{schachter1962cognitive}, despite its historical importance, has been quite controversial. Critics have argued that the use of epinephrine seems to cause different physiological reactions among subjects and may not be a reliable manipulation \cite{plutchik1967critique}. The magnitude of the effects in the Schachter \& Singer (1962) study is small and some are not statistically significant, as subsequent replicating studies yield varying success. In their replication study, Marshall \& Zimbardo \citep{marshall1979affective} found that the behavior of the confederate has little influence on the subjects. Another replication study by Maslach \citep{maslach1979negative} used hypnotic suggestions to cause the state of arousal instead of using epinephrine, and found uniformly negative experienced emotion in all groups. Finally, the study of Erdmann  \&  Janke \citep{erdmann1978interaction} used oral administration instead of injection of ephedrine to induce arousal, and added
an anxiety condition in addition to the euphoric and anger condition in the Schachter \& Singer (1962) study. In the anxiety condition, the subjects were told they would receive electric shocks, and were given mild shocks. Results show that the reported emotion from the euphoric and angry conditions conforms with the two-factor theory but increased arousal did not affect the state of anxiety among subjects.

\subsubsection{Misattribution of arousal paradigm}

Despite the criticism of the Schachter \& Singer (1962) study, it inspired a large body of subsequent research under an alternative "Misattribution of Arousal" paradigm, which provided strong support for the Two-Factor theory \cite{cotton1981review}. The Misattribution of Arousal paradigm refers to the phenomenon where individuals attribute physiological arousal to an incorrect source. An example is Ross et al. \citep{ross1969toward}, which used the misattribution effect to induce fear reduction. The participants were recruited for a learning task to solve a puzzle. If they failed to solve the puzzle, they would receive an electric shock. During the experiment, some background noise was present. When the subjects were told that their fear-related symptoms were due to effects of the background noise, the subjects reported less fear than those who were informed otherwise. In other words, when the experimenter manipulated the subjects' cognitive appraisal, the subjects misattributed their bodily state to fear-neutral sources, causing fear reduction.

\subsection{Bayesian Inference on Emotion Process}

Previous work has applied the Bayesian framework to emotion appraisal, i.e., to the mental models of an agent's emotional states. Baker et al. \citep{baker2017rational} propose a Bayesian Theory of Mind (BToM) model of how humans infer other agents' mental states, such as desires and beliefs, which can be applied to formalize emotion concepts \cite{saxe2017formalizing}. As emotions can cause agents' to display certain behaviors and expressions, an observer can infer others' underlying emotions from observations. The model of \citep{ong2019computational} deals with a variety of tasks such as attributing emotional reactions and reasoning about others’ emotion from multiple emotional cues, with rich causal linkage between emotion and events. 

These various Bayesian frameworks of emotion inference are centered on emotion appraisals on other agents rather than a direct, subjective perception of one's own emotional state. Our proposed model is hence a very different kind -- we attempt to draw a parallel between visual perception and emotion perception by building on the Two-Factor theory and modeling emotion as a Bayesian inference through which the subject appraises their {\it own} physiological arousal and contexts to produce an emotion label.

\subsubsection{Drift-Diffusion Model}
DDM is popularly used for dynamic information accumulation during perception and decision-making \citet{bitzer2014perceptual,bogacz2006physics}. In a DDM, which is often implemented with random walk process, the decision-maker accumulates evidence until the relative decision value meets one of the two decision boundaries, and a choice is made corresponding to the boundary being crossed; the corresponding choice is then selected to be the resulting decision \cite{fudenberg2020testing}. Starting point (initial bias), boundary separation, and drift rate are parameters of the DDM. Boundary separation is effectively manipulated by changing the step size, and boundary shifts effectively by moving the initial bias. In a DDM simulation, Response Time (RT) refers to the first-exit time of the drift process crossing the predetermined boundary.

Mathematically, the relative decision value $Z_t$ at any given time $t$ is modeled by 
\begin{equation}
    Z_t=Z_{t-1}+d\cdot B_t
\end{equation}
where $B_t$ is a standard Brownian motion and $d$ is the step size. The boundary crossing (first-exit) time $\tau$ is
\begin{equation}
    \tau=\inf_{}\{t\geq 0:|Z_t|\geq1\}
\end{equation}

\section{Study 1: The Schachter \& Singer Experiment}

In the original Schachter \& Singer (1962) study, the subjects are injected with either epinephrine or saline solution as a placebo. The epinephrine injection would increase the state of arousal, which is usually accompanied by tremor, palpitation, and increased heart rate. The experimental conditions of the Schachter \& Singer (1962) study are as follows with respect to Epinephrine manipulation: the {\it Informed} (Epi Inf) group receives an Epinephrine injection and is informed of the true effect of Epinephrine injection. The {\it Ignorant} group is injected with Epinephrine but is not informed of any information about the effect of the injection. The {\it Misinformed} group is injected with Epinephrine but receives incorrect information about the effect of the injection, such as numbness, itching and slight headache. The {\it Placebo} group receives the saline solution and is not informed of any information about the effect of the injection. The subjects then interact with a confederate. The confederate in the euphoric condition performs positive acts and the one in the anger condition performs negative acts. After the interaction with the confederate, the participants then complete a 9-point emotion scale. The scale has two items, measuring the feelings of anger and happiness, respectively. The emotion score is calculated by subtracting the score on the happiness item by the one on the anger item. The final emotion score ranges from extremely angry (-4) to extremely happy (4).

The results of the study are re-plotted in Tables 1 and 2. Schachter \& Singer (1962) found that, in the euphoric condition where the confederate performed positive acts, the mean self-reported emotion scores of the Epi Ignorant and Epi Misinformed groups were greater than the mean of the Informed group. In other words, in the condition when the confederate is euphoric, subjects reported to be more euphoric when they had no explanation of their own bodily states than when they did. A similar pattern is observed in the anger condition where the confederate performed negative acts. 

\begin{table}[h]

\vspace*{-0.3cm}
\caption{Emotion Scores in the Euphoric State}
\vspace{0.1cm}
\begin{center} {\footnotesize
\begin{tabular}{ccc}
\hline 
  \multicolumn{1}{c}{Experimental Conditions} & \multicolumn{1}{c}{N} & \multicolumn{1}{c}{Self-reported Scale} \\
\hline
Epi Informed & 25 & 0.98 \\
Epi Misinformed & 25 & 1.78 \\
Epi Ignorant & 25 & 1.90 \\
Placebo & 26 & 1.61 \\

\hline
\end{tabular} }
\end{center}

\label{turns}
\end{table}

\begin{table}[h]
\vspace*{-1cm}
\caption{Emotion Scores in the Anger State}
\vspace{0.1cm}
\begin{center} {\footnotesize
\begin{tabular}{ccc}
\hline 
  \multicolumn{1}{c}{Experimental Conditions} & \multicolumn{1}{c}{N} & \multicolumn{1}{c}{Self-reported Scale} \\
\hline
Epi Informed & 22 & 1.91 \\
Epi Ignorant & 23 & 1.39 \\
Placebo & 23 & 1.63\\

\hline
\end{tabular} }
\end{center}

\label{turns}
\vspace*{-0.5cm}
\end{table}

\subsection{Methods}

We identify the response times of a DDM simulation to self-reported emotion rating in the Schachter \& Singer (1962) experiment, and assume that a longer response time indicates a less intense emotion (as the individual needs more time to accumulate enough emotion-related evidence to reach the threshold). To map the physiological arousal factor and cognitive appraisal factor onto the prior and the likelihood factors, which corresponds to the drift rate and the initial bias in the DDM model, we investigated two model scenarios (Table 3). In Model 1, cognitive appraisal of the context is modeled by drift rate, as the interaction with the confederate provides accumulating evidence in deciding one's emotion state. The Epi Informed group has the lowest drift rate ($v_0<v_1$) because the arousal is explained and thus the context becomes less relevant. In Model 2, cognitive appraisal of the context is modeled by initial bias; here the Epi Informed group should have the lowest initial bias ($s_0<s_1$) because the arousal is already explained and the context is less relevant to begin with. In both scenarios, the placebo group has a smaller step size ($d_0>d_1$), which models the state of arousal.

We measure the model goodness of fit with Mean Squared Error (MSE), which measures the average of the squares of the errors when comparing the model predictions and the experimental results.

\begin{table*}[h]
\vspace*{-0.5cm}
\caption{Parameter settings for the DDM}
\vspace{0.1cm}
\begin{center} {\footnotesize
\begin{tabular}{ccccccccc}
\hline 
  \multicolumn{1}{c}{Conditions} & & \multicolumn{3}{c}{Model 1} &  & \multicolumn{3}{c}{Model 2} \\\cline{3-5}  \cline{7-9}
  & &Initial Bias & Drift Rate & {Step Size}   & & {Initial Bias} & {Drift Rate} &{Step Size} \\
\hline
Epi Informed & &0 & $v_0$ &$d_0$ && $s_0$&0 &$d_0$ \\
Epi Misinformed & &0 & $v_1$ &$d_0$&&$s_1$&0 &$d_0$\\
Epi Ignorant & &0 & $v_1$ &$d_0$&&$s_1$&0&$d_0$ \\
Placebo & & 0 & $v_1$ &$d_1$&&$s_1$&0 &$d_1$\\

\hline
\end{tabular} }
\end{center}
\label{turns}
\vspace*{-0.5cm}
\end{table*}

\begin{figure}
    \centering
    \includegraphics[width=4cm]{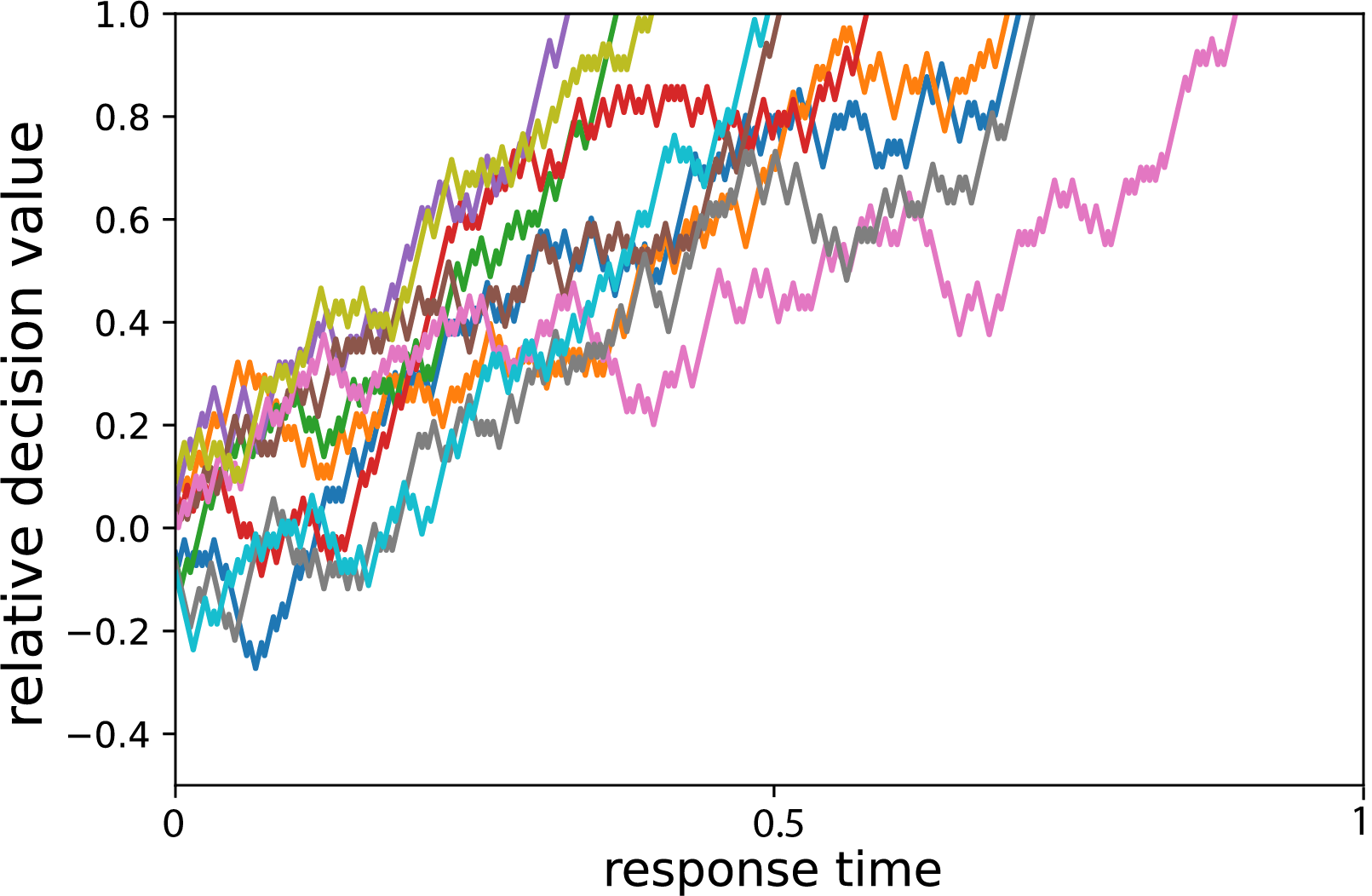}
    \includegraphics[width=4cm]{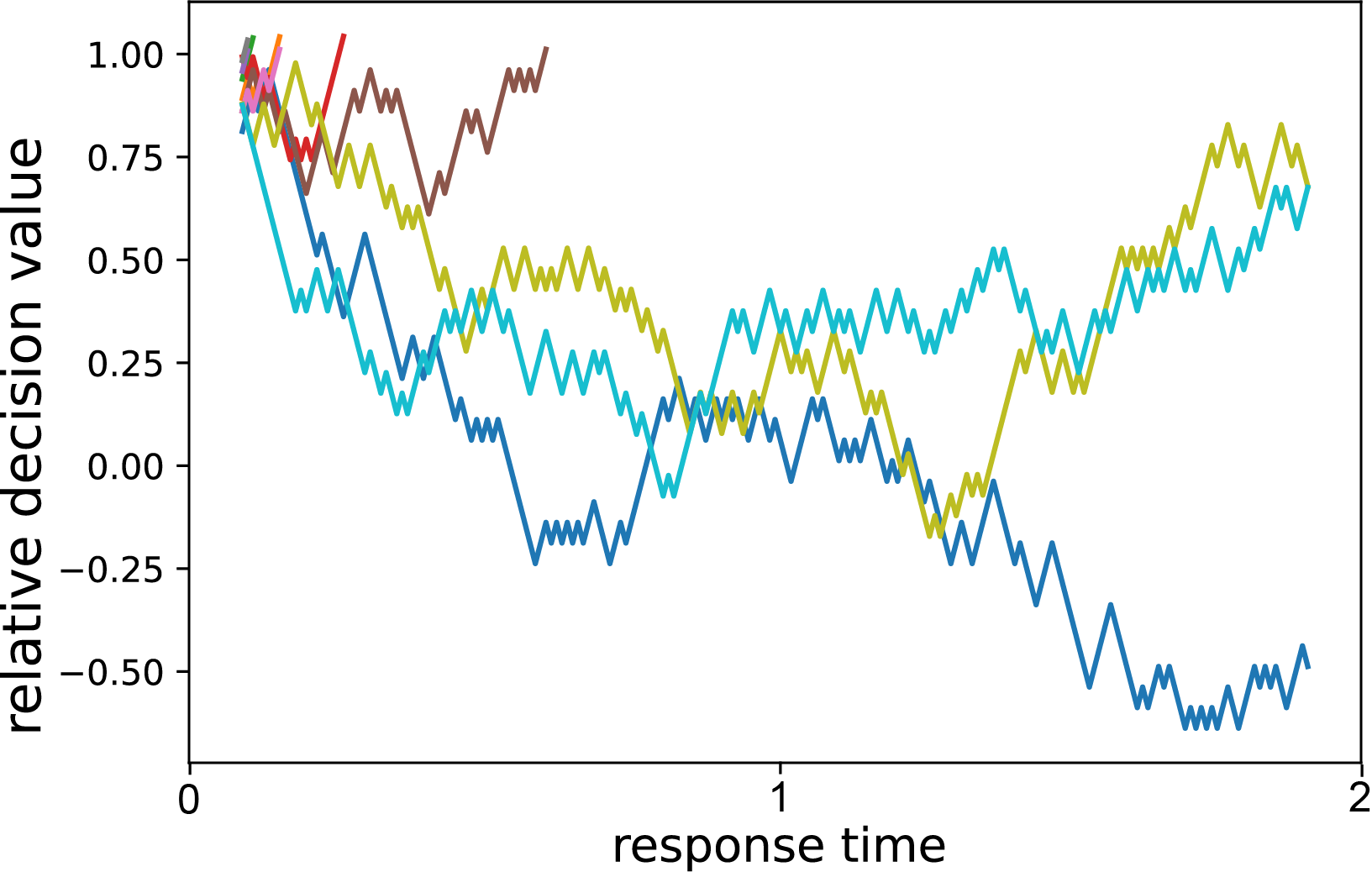}
    \caption{Sample simulation paths of Model 1 (left) and Model 2 (right)}
    \label{fig:my_label}
    \vspace*{-0.3cm}
\end{figure}

Model parameters are given in Table 3. A grid search was used to find the best-fitting parameters. Each experimental condition was simulated with 1000 trials. 
After simulating the results of the 4 experimental conditions, we normalized the results to ensure the simulated ratings had the same aggregated value as the experimental results.

\subsection{Results}
Sample paths with 10 trials are shown in Figure 1. For Model 1 (arousal as the prior and context as the likelihood function), paths are stable with little variations, with trajectories showing a steady evidence accumulation process. For Model 2 (context as the prior and arousal as the likelihood function), paths are divergent with greater variations in response time, with some crossing the wrong decision boundaries ("error trials").

Both Model 1 and Model 2 show a similar trend as the experimental results (Fig 2), where the Epi Ignorant and Epi Informed group have the highest reported value in the euphoria condition, followed by the Placebo group and the Epi Informed group. In the anger condition, the Epi Informed group has the highest emotion value, followed by the placebo group and the Epi Ignorant group. 

Model 1 (MSE = 0.0072) seems to have a better fit with the experimental results than Model 2 (MSE = 0.1852), judging from the match in emotion scores and their variance for each experimental group. We conclude that arousal is better modeled as the prior and context as a likelihood function.

\begin{figure*}

    \centering
    \includegraphics[width=6.5cm]{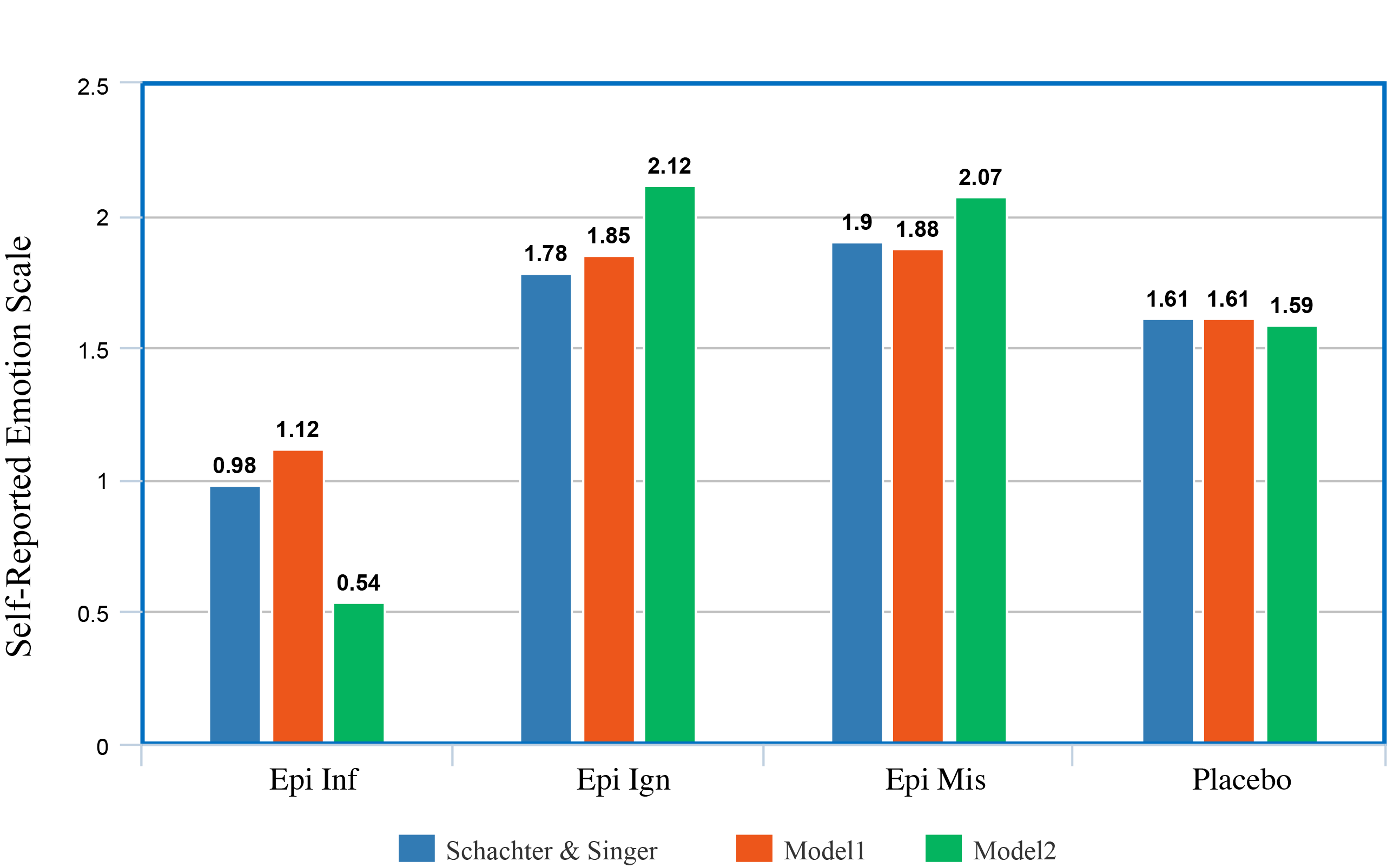}
    \includegraphics[width=6.5cm]{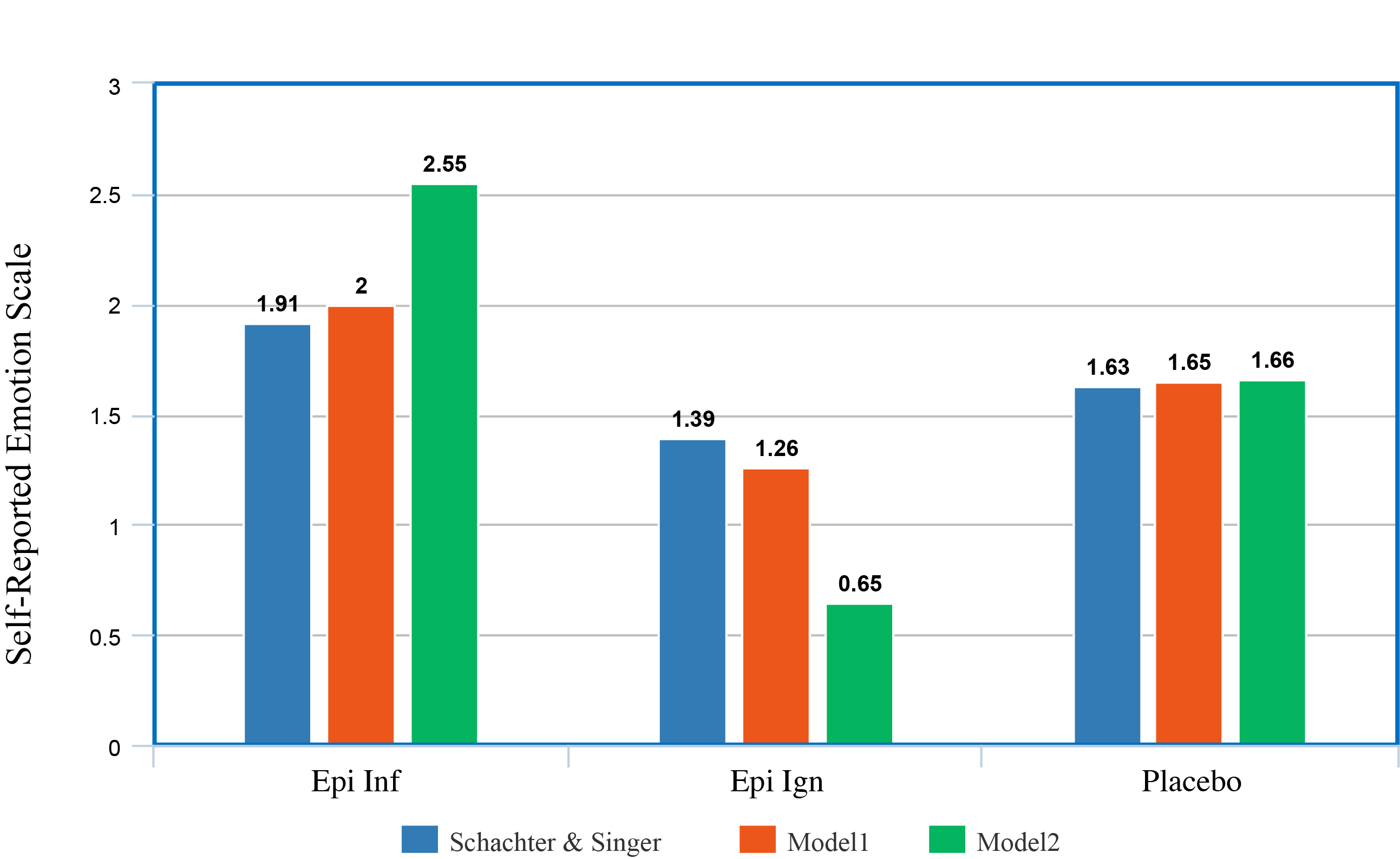}
    \label{fig:my_label}
    \caption{Experimental and simulation results for the Euphoria State (left) and Anger State (right)}
    \vspace*{-0.5cm}
\end{figure*}

\section{Study 2: Time-Course in Misattribution Effect}

In Ross et al. (1969), misattribution of arousal successfully led to the reduction of fear of pending electric shocks. In the experiments, 40 subjects are recruited to solve two puzzles, a "reward" puzzle and a "shock" puzzle. The experimenters first played a loud, unpleasant noise for 3 minutes, after which the subjects have 3 minutes to work on the puzzles. Solving the reward puzzle lead to monetary rewards while solving the shock puzzle enable participants to avoid an electric shock as punishment. In the experiment, the anticipation of an electric shock presents as a stimulus for fear.  Subjects are randomly assigned into two experimental groups: shock-attribution group and noise-attribution group, and the amount of time each subject spends on either of the two puzzles is recorded. 

Both groups are briefed about likely symptoms caused by the noise before the experiment begins. In the {\it shock-attribution} group, the subjects are informed that they may experience with symptoms related to noise bombardment, such as numbness or dizziness and headaches, whereas in the {\it noise-attribution} group, the subjects are informed that they may experience fear-related symptoms, such as tremor, palpitations, etc. Subjects in the noise-attribution (i.e., misattribution) group are expected to be less fearful as they are more likely to attribute their arousal to the presence of noise than the pending electric shock. The average time participants spend on the shock-avoidance puzzle is used as a surrogate of their fear level, since subjects more fearful of the electric shock are likely to spend more time avoiding the shock than getting monetary rewards. The percentage of subjects working on the shock-avoidance puzzle in each group is recorded every 10 seconds.

Ross et al. (1969) found that the noise-attribution group spent significantly less time on the shock-avoidance puzzle (M = 88.22, SD = 57.02) than the shock-attribution group (M = 140.34, SD = 34.84). The percentage of participants working on the shock-avoidance puzzle over 10s intervals is shown in Fig. 3. Most subjects in both groups start focusing on the shock-avoidance puzzle. While the percentage of subjects working on the shock-avoidance puzzle remains at around 80\% in the shock-attribution group, more subjects in the noise-attribution group start working on the reward puzzle as time goes by.


\begin{figure}

    \centering
    \includegraphics[width=7cm]{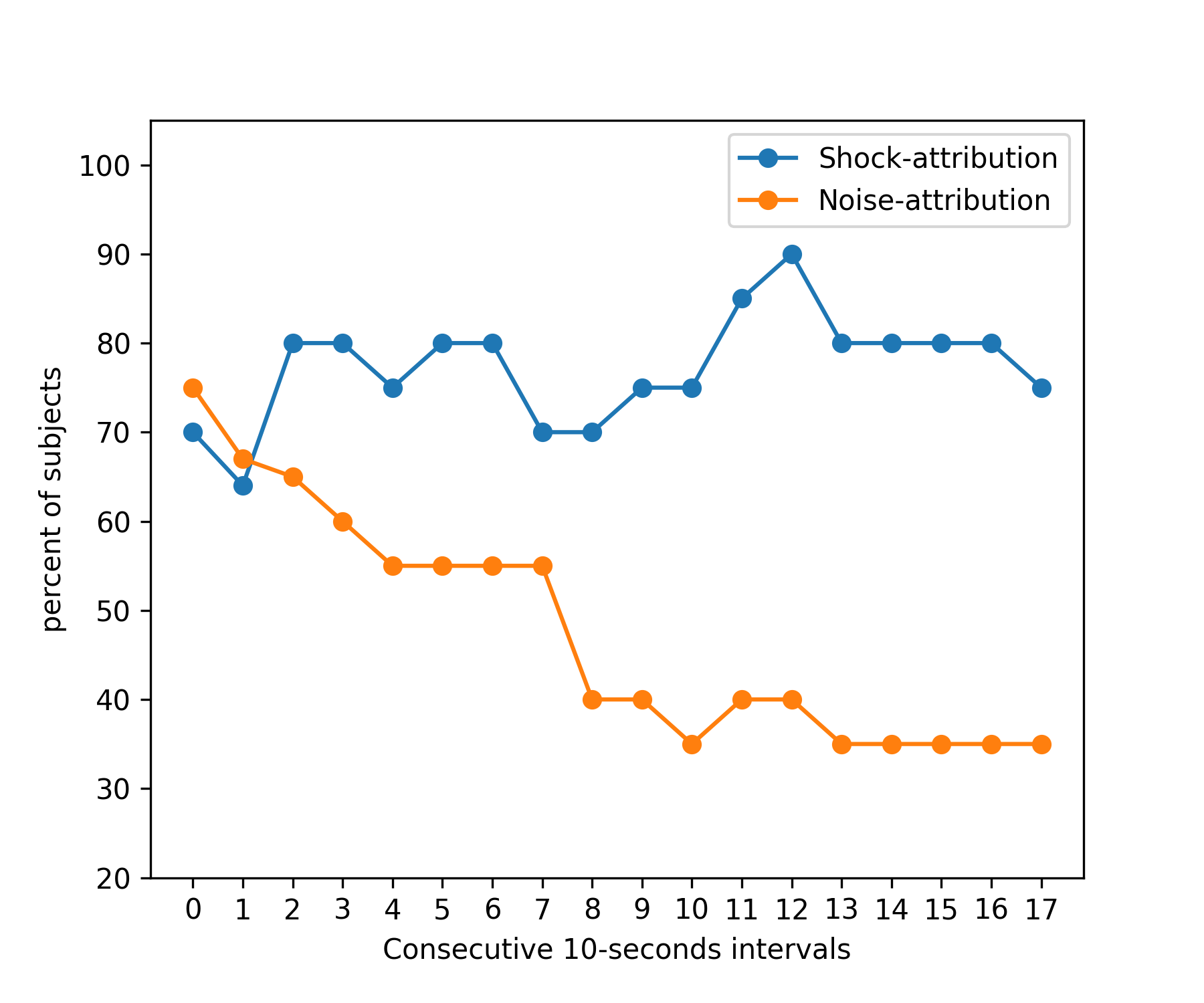}
    \label{fig:my_label}
    \caption{The percentage of subjects in each condition working on the shock-avoidance puzzle during each 10-sec time interval (data from Ross et al. (1969) replotted).}
    \vspace*{-0.5cm}
\end{figure}

\begin{table*}[h]
\caption{Parameter settings for the DDM for the t-th time interval}
\vspace{0.1cm}
\begin{center} {\footnotesize
\begin{tabular}{ccccccccc}
\hline 
  \multicolumn{1}{c}{Conditions} & & \multicolumn{3}{c}{Model 1} &  & \multicolumn{3}{c}{Model 2} \\\cline{3-5}  \cline{7-9}
  & &Initial Bias & Drift Rate & Boundaries  & & {Initial Bias} & {Drift Rate} & Boundaries \\
\hline
Shock Attribution & &$s_0$ & $v_0$ & $\pm1$  && $s_0$&$v_0$ & $\pm1$\\
Noise Attribution & &$s_0$ & $v_0-\alpha_t$ &$\pm1$ && $s_0$&$v_0$ & $\pm1-\alpha_t$\\

\hline
\end{tabular} }
\end{center}
\label{turns}
\vspace*{-0.5cm}
\end{table*}

\subsection{Methods}

To simulate the experimental paradigm, we assign the boundaries of the DDM to be fear (positive boundary) and non-fear (negative boundary). Similar to Study 1, we explored two model scenarios: Model 1 uses the arousal as the prior and the context as the likelihood; Model 2 uses the context as prior and the arousal as the likelihood. In contrast to the Schachter \& Singer (1962) study, the initial bias and drift rate are both set as positive in both model groups since anticipation of fear is present and associated with a high physiological arousal state. To model the misattribution effect, we assign a time-dependent variable $\alpha$ to model either the decrease in arousal (Model 1) or decrease in boundary separation (Model 2). We assume that as the subjects increasingly attribute their arousal to the emotion-neutral source, they reduce the association between their bodily state and the emotion of fear. This interpretation is consistent with previous work on the appraisal theory, where cognitive appraisal can be used to reduce arousal \cite{lazarus1964short,dandoy1990use}. In Model 1 where the arousal is the prior, the drift rate decreases by to $v_0-\alpha_t$ at each time interval. In Model 2 where context is the prior, we model the time-dependent changes in arousal appraisal by shifts in the decision boundaries, where the positive and negative boundary changes to $1-\alpha_t$ and $-1-\alpha_t$, respectively. The time-dependent shift in boundary is consistent with the notion of ``collapsing boundary'' in DDM models used for decision making under time pressure \citet{palestro2018some, hawkins2015revisiting}. The parameters for the two models are shown in Table 4. 
\begin{figure}

    \centering
    \includegraphics[width=6cm]{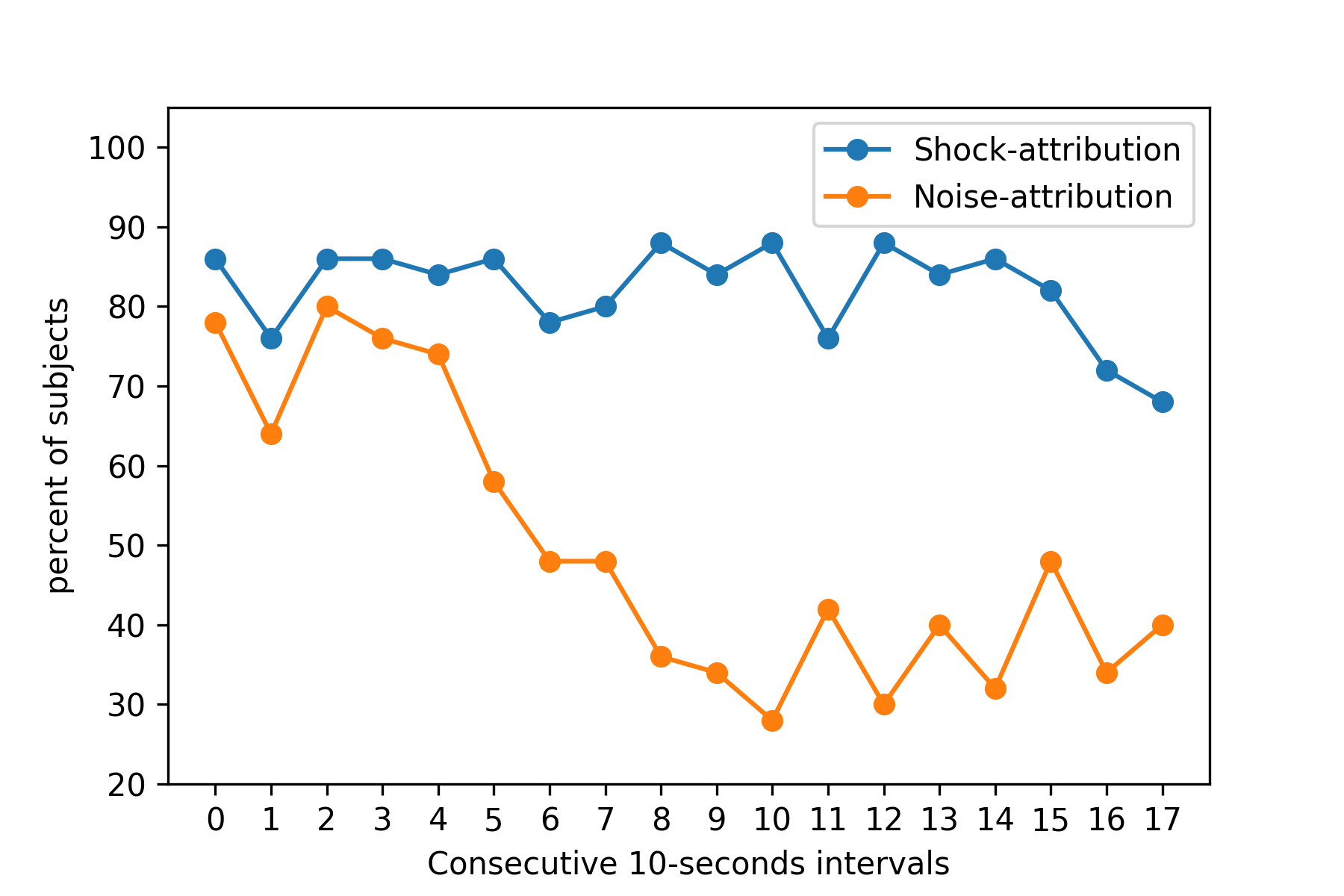}
    \includegraphics[width=6cm]{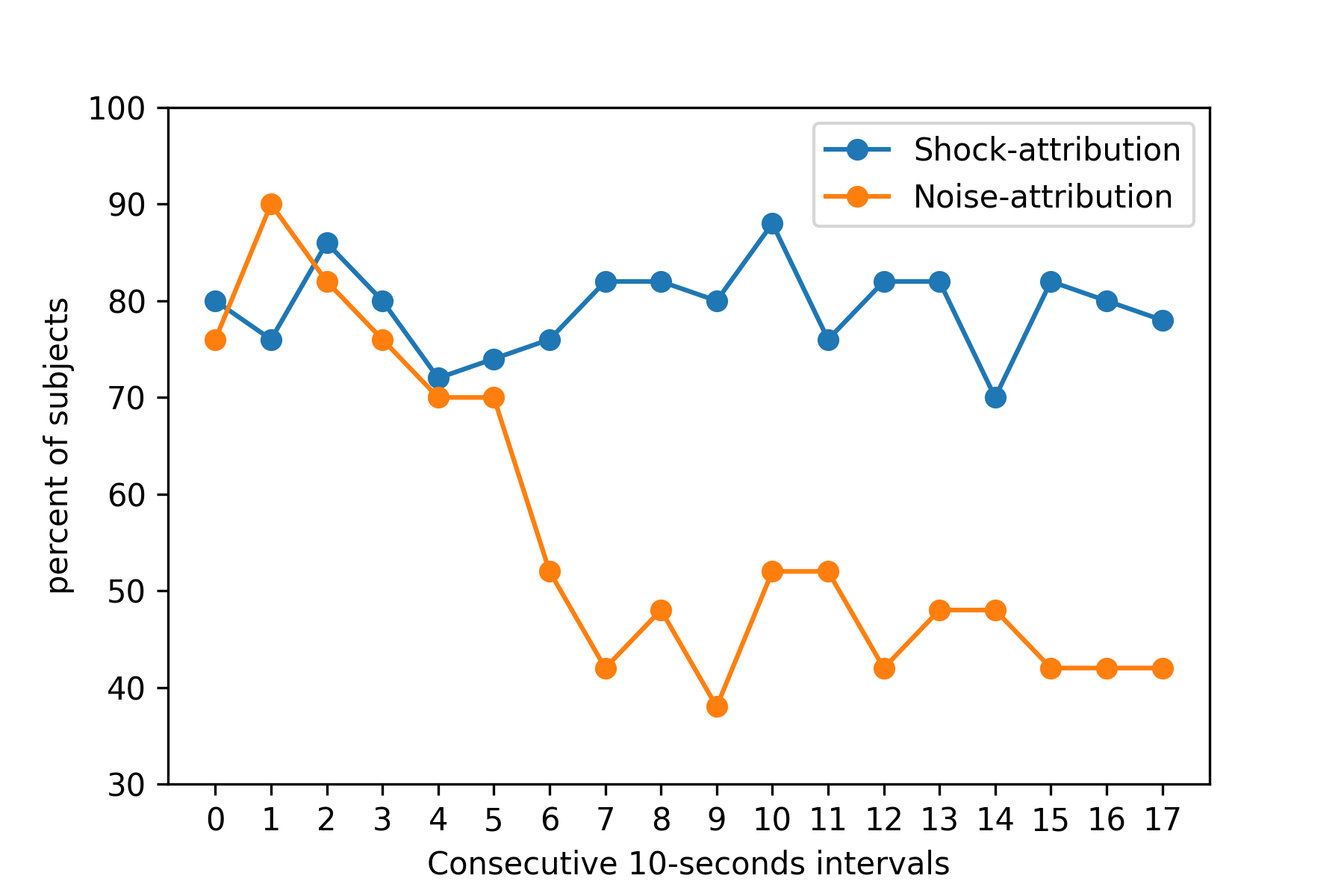}
    \caption{Simulated results of Model 1 (top) and Model 2 (bottom)}
    \label{fig:my_label}    
    \vspace{-0.5cm}
\end{figure}

For each time interval, we run a simulation with 50 trials for each experimental condition. Similar to study 1, we measure the intensity of fear by the response time of the DDM. We map the response time to a 5-point fear scale, with 0 indicating not fearful at all and 4 indicating extremely fearful. Crossing the non-fear boundary would correspond to a score of 0, regardless of the response time. With a step size of 0.05, the simulation needs at least 20 steps to cross the fear boundary. Therefore, a response time of 20 would correspond to extreme fearfulness on the scale. 

A fear score over 2 (midpoint of the 5-point fear scale) is taken to mean that the subject is assumed to be working on the shock-avoidance puzzle at a given time point. Therefore, the proportion of trials (out of all 50 trials) with a fear score over 2 is a proxy for the percentage of subjects working on the shock-avoidance puzzle.

Because the data is aggregated across subjects, the decision boundaries can be seen as the threshold probability, where the relative decision value is the posterior probability differential between the two alternatives at time t, and prior probability at time t+1. At a given time t, the posterior probability increases or decreases by a given step size with the likelihood function modeled as the drift rate.

\subsection{Results}

The simulation results are shown in Figure 4. Both models are able to show a similar trend as the Ross et al. (1969) experimental data where both experimental conditions start with a high percentage of participants working on the shock-avoidance puzzle. The percentage in the noise-attribution group keeps decreasing till the 7th-10th interval and fluctuates at around 40\%. 
In the noise-attribution group, the trajectory after the 8th interval tends to have less variance in Model 2 (SD = 4.64) as opposed to in Model 1 (SD = 5.73). But otherwise, the two Models are quite comparable.  

\section{Discussions and Conclusion}

We formulated a Bayesian framework of Schachter \& Singer's Two-Factor theory of emotion, where the physiological arousal and cognitive label factors are mapped onto the prior or the likelihood function factors of a Bayesian framework (and initial bias or drift-rate of a Drift-Diffusion Model). Parameterization of DDMs is flexible enough to model the attribution process crucial for emotion as reflected in the experimental data of both the Schachter \& Singer (1962) and the Ross et al. (1969) study. 

We explored how physiological arousal and appraisal of contextual information are mapped differently onto the prior and the likelihood. Model 1 (arousal as the prior and context as the likelihood) shows a better fit for the Schachter \& Singer (1962) study, but not for the Ross et al.\ (1969) study. This may be due to a difference in the experimental paradigms. In the Schachter \& Singer (1962) study, there is continuous interaction between the participants and the confederate, and hence a continuing target for cognitive appraisal (evidence accumulation), whereas in the Ross et al.\ (1969) study, the emotion-related stimulus noise is presented only before participants work on the puzzle, while electric shock comes only after the trial ends. Rather, there is the deadline aspect (as opposed to evidence accumulation aspect) in this paradigm. Another possible difference is that the Schachter \& Singer (1962) study measures both the polarity of the emotional states (euphoric and anger) and the intensity whereas, in the Ross et al. (1969) study, only the level of fear is approximated. So arousal is unbiased and neutral in the former, but not in the latter. 

Due to the lack of individual experimental data, we are not able to fully assess the goodness of fit of our simulations. Though the two proposed model setups in our experiments show different behaviors, we are not able to make a rigorous quantitative comparison. Nevertheless, our study proposes a Bayesian model for emotion process based on Schachter \& Singer's theory of emotion and our simulation shows some encouraging preliminary results. Future studies can test the model against individual experimental data and conduct a more rigorous model evaluation.

\bibliographystyle{apacite}

\setlength{\bibleftmargin}{.125in}
\setlength{\bibindent}{-\bibleftmargin}

\bibliography{BibliographyFile}

\end{document}